\documentclass[twocolumn,pra,aps,showpacs]{revtex4}

\usepackage{mathptmx}
\usepackage{psfrag,graphicx}
\usepackage{dcolumn}
\usepackage{amsmath,amssymb,amsfonts}
\usepackage{bm}
\usepackage{latexsym}
\usepackage{epstopdf}
\usepackage{color}
\usepackage[english]{babel}
\usepackage{natbib}
\usepackage{appendix}
\usepackage[normalem]{ulem}

\def\ba{\begin{array}}
\def\ea{\end{array}}
\def\bi{\begin{itemize}}
\def\ei{\end{itemize}}

\def\be{\begin{equation}}
\def\ee{\end{equation}}
\def\bea{\begin{eqnarray}}
\def\eea{\end{eqnarray}}
\def\bse{\begin{subequations}}
\def\ese{\end{subequations}}
\def\bt{\begin{tabular}}
\def\et{\end{tabular}}
\def\bc{\begin{center}}
\def\ec{\end{center}}

\def\tsum{\displaystyle{\sum}}
\def\tsum{\sum}

\def\hes{A}

\def\etal{\textit{et al.}}

\def\PG{P}
\def\TG{\Phi}
\def\T{\Phi}

\def\Ggate#1{G{#1}}
\def\Ngate#1{N{#1}}
\def\Ugate#1{M{#1}}
\def\GGgate#1{G{#1}}
\def\Bgate#1{B{#1}}

\def\adr{\affiliation{Department of Physics, St. Kliment Ohridski University of Sofia, 5 James Bourchier Blvd, 1164 Sofia, Bulgaria}}

\def\ccc{three-qubit conditional phase}
\def\cccs{three-qubit c-phase}

\def\tsum{\sum\nolimits}
\def\id{\mathbf{1}}

\newcommand{\bra}[1]{{\left\langle{#1}\right\vert}}
\newcommand{\ket}[1]{{\left\vert{#1}\right\rangle}}

\begin{document}

\author{Svetoslav S. Ivanov}
\adr
\author{Peter A. Ivanov}
\adr
\author{Nikolay V. Vitanov}
\adr
\title{Efficient construction of three- and four-qubit quantum gates by global entangling gates}

\begin{abstract}
We present improved circuits for the control-control-phase (Toffoli) gate and the control-swap (Fredkin) gate using three and four global two-qubit gates, respectively. This is a nearly double speed-up compared to the conventional circuits, which require five (for Toffoli) and seven (for Fredkin) conditional two-qubit gates. We apply the same approach to construct the conditional four-qubit phase gate by seven global two-qubit gates. We also present construction of the Toffoli and the Fredkin gates with five nearest-neighbour interactions. Our constructions do not employ ancilla qubits or ancilla internal states and are particularly well suited for ion qubits and for circuit QED systems, where the entangling operations can be implemented by global addressing.
\end{abstract}

\maketitle

%%%%%%%%%%%%%%%%%%%%%%%%%%%%%%%%%%%%%%%%%%%%%%%%%%%%%%%%%%%%%%%%%%%%%%%%%%%
%%%%%%%%%%%%%%%%%%%%%%%%%%%%%%%%%%%%%%%%%%%%%%%%%%%%%%%%%%%%%%%%%%%%%%%%%%%
%%%%%%%%%%%%%%%%%%%%%%%%%%%%%%%%%%%%%%%%%%%%%%%%%%%%%%%%%%%%%%%%%%%%%%%%%%%
\section{Introduction}\label{Sec-introduction}
%%%%%%%%%%%%%%%%%%%%%%%%%%%%%%%%%%%%%%%%%%%%%%%%%%%%%%%%%%%%%%%%%%%%%%%%%%%%%%%%%%%%%%%%%%%%%%%%%%%%%%%%%%%%%%%%%%%%%%%
The three-qubit Toffoli and Fredkin gates are
arguably the most important highly-conditional quantum gates in quantum information science
 because they are the key enabling ingredient in quantum error correction \cite{Cory1998,Ekert1996}.
It is known that these gates are universal: any quantum computation can be constructed entirely by Fredkin or Toffoli gates, combined with the one-qubit Hadamard and phase gates \cite{Shi,Nielsen2000}.
The Toffoli gate or the closely related control-control-phase gate (cc-phase gate) has been experimentally demonstrated with nuclear magnetic resonance (NMR) \cite{Cory1998},
linear optics \cite{Lanyon2009,Micuda2013}, trapped ions \cite{Monz09},
and recently with superconducting qubits \cite{Mariantoni2011,Reed2012,Fedorov2012}. The Fredkin gate has been demonstrated with NMR \cite{Fei2002}.

The standard decomposition of the Toffoli gate in the circuit model of quantum computation uses six CNOT (or c-phase) gates \cite{Barenco1995,Nielsen2000}.
It can be constructed also with five conditional two-qubit gates --- two CNOT gates and three controlled-V gates, where V is the $\sqrt{\text{NOT}}$ gate \cite{Barenco1995,Jones98} (see Appendix \ref{Appendix:notation}).
The Fredkin gate can be obtained from the Toffoli gate if surrounded with two CNOT gates \cite{Smolin96}.
Recently, it was shown that five conditional two-qubit gates are not only sufficient but also necessary for the implementation of the Toffoli gate \cite{Yu2013} and the Fredkin gate \cite{Smolin96,Yu2013Arxiv}.

The speed of quantum circuits is usually determined by the number of conditional gates and the objective is to minimize this number.
Over the last years, several alternative methods for more efficient construction of the three-qubit gates have been proposed and demonstrated.
Following a theoretical proposal by Ralph \etal\ \cite{Ralph2007}, Lanyon \etal\ \cite{Lanyon2009} constructed the Toffoli gate in a linear optics experiment with only three two-qubit gates by using a qutrit target, i.e., with an ancilla state added to the target qubit.
Ivanov and Vitanov \cite{Ivanov2011} proposed a method for constructing highly-conditional C$^n$-NOT gates,
 which uses composite pulses, without two-qubit gates.
Monz \etal\ \cite{Monz09} have demonstrated experimentally the Toffoli gate with
calcium ions in a linear Paul trap by using a sequence of 15 laser pulses;
13 of these pulses were on the first blue-sideband transition and they performed reversible mapping between electronic and vibrational motion, as in the Cirac-Zoller's CNOT gate \cite{Cirac95,Monroe95,Schmidt-Kaler03}.

In this paper, we propose realizations of the three- and the four-qubit conditional-phase gates and the Fredkin gate using global entangling gates and no ancilla internal states or ancilla qubits. Two types of physical systems are considered for the realization of the entangling gates, where i) all qubits interact simultaneously (global interaction), or where ii) each qubit interacts with its nearest neighbours only (nearest-neighbour interaction).
The three- and the four-qubit phase gates are constructed with only three and seven global interactions, respectively, and the Fredkin gate -- with four.
Each of the global two-qubit gates is implemented in a single step, by global addressing of all qubits, which amounts to simultaneous two-qubit phase gates upon each pair of qubits.
These realizations are particularly well suited for trapped ions in a laser-driven linear Paul trap \cite{Wineland,Haffner,Wineland-Nobel}, and in a microwave-driven trap with magnetic-field gradient \cite{Mintert2001,Timoney2011,Webster2013,Johanning2009}.
We note that this speed-up does not contradict the earlier findings that five two-qubit gates are both necessary and sufficient for the three-qubit Toffoli and Fredkin gates \cite{Smolin96,Yu2013,Yu2013Arxiv}, because each global gate used here embraces three two-qubit gates. The total number of two-qubit gates is therefore larger than five; however, they are conveniently grouped into fewer global gates.

With nearest-neighbour interactions the three-qubit c-phase and Fredkin gates require five entangling gates.
These realizations are suitable, e.g., for circuit cavity QED \cite{Blais}, where the nearest-neighbour coupling arises naturally from the hopping of photons between adjacent cavities in the regime of photon blockade \cite{Illuminati}.

Global addressing is the key tool in the S{\o}rensen-M{\o}lmer two-qubit c-phase gate \cite{Sorensen} and in their proposal for generation of GHZ states \cite{Molmer}.
It has been used recently in the construction of the Toffoli gate by only three global conditional operations~\cite{Nebendahl2009}.
Here we extend this method to other gates and also present constructions with global addressing in the important case of nearest-neighbor interaction.

Physical realizations with trapped ions are also discussed.
In the laser-driven Paul trap we use bichromatic laser fields, the two components of which are tuned at a certain detuning from the first red and blue sidebands of the qubit transition.
These fields address simultaneously all qubits at each step.
With suitably chosen laser phases these bichromatic fields create effective spin-spin interactions between the qubits and cancel the dependence on the phonon number.
The thereby designed Toffoli and Fredkin gates are thus both faster and applicable to ions in a thermal state of motion, like the S{\o}rensen-M{\o}lmer c-phase gate \cite{Sorensen}.

In the microwave-driven magnetic-gradient trap \cite{Mintert2001,Timoney2011,Webster2013,Johanning2009}, the gates are implemented in a particularly simple fashion because the required spin-spin couplings arise naturally from the magnetic-field gradient.
The conditional two-qubit gates are implemented merely by letting the system evolve freely for a certain time duration.

%%%%%%%%%%%%%%%%%%%%%%%%%%%%%%%%%%%%%%%%%%%%%%%%%%%%%%%%%%%%%%%%%%%%%%%%%%%%%%%%%%%%%%%%%%%%%%%%%%%%%%%%%%%%%%%%%%%%%%%
\section{Fredkin, Toffoli and \cccs~gates: Mathematical construction}\label{Sec-theory}
%%%%%%%%%%%%%%%%%%%%%%%%%%%%%%%%%%%%%%%%%%%%%%%%%%%%%%%%%%%%%%%%%%%%%%%%%%%%%%%%%%%%%%%%%%%%%%%%%%%%%%%%%%%%%%%%%%%%%%%

In the Toffoli gate, there are two control qubits and a target qubit: the target qubit is inverted if the two control qubits are in state $\ket{1}$, but it is left unchanged otherwise.
In the three-qubit basis,
\be
\label{basis}
\{\ket{000},\ket{001},\ket{010},\ket{011}, \ket{100}, \ket{101},\ket{110},\ket{111}\},
\ee
the Toffoli gate has the matrix form
\be
\bt{ccc}
\bt{c}\includegraphics[width=0.08\columnwidth]{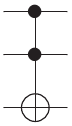}\et
& \bt{c} = \et &
$\left[\begin{array}{cccccccc}
1 & 0 & 0 & 0 & 0 & 0 & 0 & 0 \\
0 & 1 & 0 & 0 & 0 & 0 & 0 & 0 \\
0 & 0 & 1 & 0 & 0 & 0 & 0 & 0 \\
0 & 0 & 0 & 1 & 0 & 0 & 0 & 0 \\
0 & 0 & 0 & 0 & 1 & 0 & 0 & 0 \\
0 & 0 & 0 & 0 & 0 & 1 & 0 & 0 \\
0 & 0 & 0 & 0 & 0 & 0 & 0 & 1 \\
0 & 0 & 0 & 0 & 0 & 0 & 1 & 0
\end{array}\right]$
\et
\ee
and the \ccc~(cc-phase) gate is
\be
\bt{ccc}
\bt{c}\includegraphics[width=0.08\columnwidth]{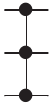}\et
& \bt{c} = \et &
$\left[\begin{array}{cccccccc}
1 & 0 & 0 & 0 & 0 & 0 & 0 & 0 \\
0 & 1 & 0 & 0 & 0 & 0 & 0 & 0 \\
0 & 0 & 1 & 0 & 0 & 0 & 0 & 0 \\
0 & 0 & 0 & 1 & 0 & 0 & 0 & 0 \\
0 & 0 & 0 & 0 & 1 & 0 & 0 & 0 \\
0 & 0 & 0 & 0 & 0 & 1 & 0 & 0 \\
0 & 0 & 0 & 0 & 0 & 0 & 1 & 0 \\
0 & 0 & 0 & 0 & 0 & 0 & 0 & -1
\end{array}\right]$.
\et
\ee
Because the cc-phase gate is symmetric, and all three qubits have the same role (no distinction of target and control gates) hereafter we shall be concerned only with it.
As it is well known, the Toffoli gate can be obtained from the cc-phase gate with two Hadamard gates,
\be
\includegraphics[width=0.4\columnwidth]{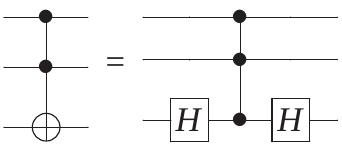}\notag
\ee

Given the cc-phase gate, any of the three qubits can be turned into a target qubit by the application of the two Hadamard gates on it.
Therefore, the implementations of the cc-phase gate below can be directly translated into Toffoli gates.

In the Fredkin gate, there is one control qubit and two target qubits: the target qubits are swapped if the control qubit is in state $\ket{1}$, but are left unchanged otherwise. In the basis \eqref{basis} the Fredkin gate has the matrix form
\be
\bt{ccc}
\bt{c}\includegraphics[width=0.08\columnwidth]{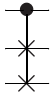}\notag\et
& \bt{c} \\ = \et &
$\left[\begin{array}{cccccccc}
1 & 0 & 0 & 0 & 0 & 0 & 0 & 0 \\
0 & 1 & 0 & 0 & 0 & 0 & 0 & 0 \\
0 & 0 & 1 & 0 & 0 & 0 & 0 & 0 \\
0 & 0 & 0 & 1 & 0 & 0 & 0 & 0 \\
0 & 0 & 0 & 0 & 1 & 0 & 0 & 0 \\
0 & 0 & 0 & 0 & 0 & 0 & 1 & 0 \\
0 & 0 & 0 & 0 & 0 & 1 & 0 & 0 \\
0 & 0 & 0 & 0 & 0 & 0 & 0 & 1
\end{array}\right]$ \label{FG}.
\et
\ee

Here we introduce improved quantum circuits, which produce the three-qubit cc-phase gate, the four-qubit ccc-phase gate and the Fredkin gate with a minimal number of conditional operations.
We consider circuits based on two types of interactions: global interactions, where all pairs of qubits interact collectively, and nearest-neighbour interactions, where each qubit interacts with its nearest neighbours only. Both realizations can be produced by fields addressing uniformly all qubits.

\subsection{Implementation with global interactions}

We begin with the design of circuits where the conditional gates are produced by global interactions, which we call \textit{G}-gates.

\subsubsection{Toffoli and cc-phase gates}\label{Sec-three}

\begin{figure}
\label{global-gates}
\includegraphics[width=\columnwidth]{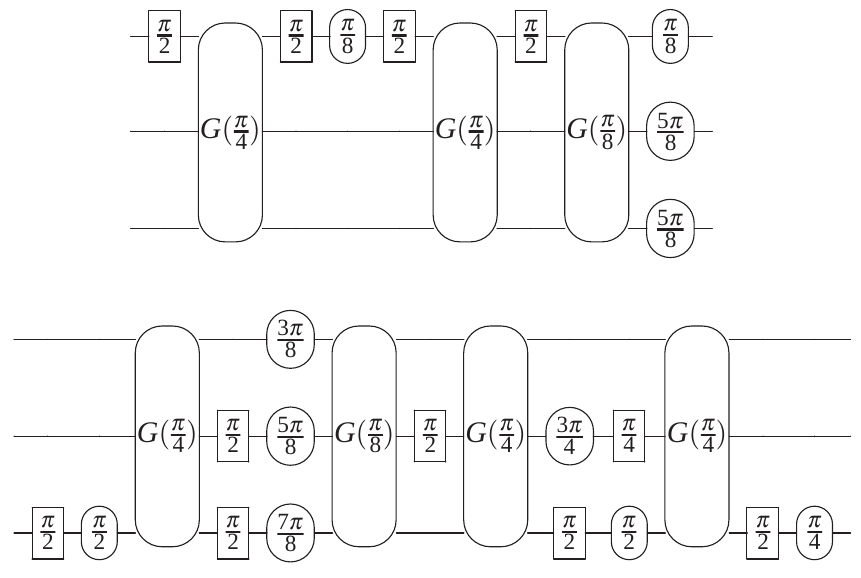}
\label{fig4}
\caption{Global realization of the three-qubit c-phase gate (top) and the Fredkin gate (bottom).}
\end{figure}

By using the method outlined in  Appendix \ref{Appendix:numerics}, we have constructed a quantum circuit of the \ccc~gate with three $\Ggate{}$-gates only, which can be verified by direct inspection. The circuit is shown in Fig. \ref{global-gates} (top). Besides the $\Ggate{}$-gates, it consists of $\pi/2$ pulses (boxes), and phase gates (ovals). See Appendix \ref{Appendix:notation} for the complete list of notations.
In operator form (to be read from left to right), the circuit reads
\be
P_1 \Ggate{(\tfrac\pi4)} P_1 \T_1(\tfrac\pi8) P_1 \Ggate{(\tfrac\pi4)} P_1 \Ggate{(\tfrac\pi8)} \T_1(\tfrac{\pi}8) \T_2(\tfrac{5\pi}8) \T_3(\tfrac{5\pi}8),
\ee
where
\bse
\begin{align}
P_k &= e^{-i \frac\pi 4 \sigma_k^x}, \\
\T_k(\phi) &= e^{-i \phi \sigma_k^z}, \\
\Ggate{(\phi)} &= e^{i \phi (\sigma_1^z \sigma_2^z + \sigma_2^z \sigma_3^z + \sigma_1^z \sigma_3^z)}. \label{B}
\end{align}
\ese
This realization demands only three conditional gates, four phase gates (in two phase groups, hence the T-depth \cite{Amy} is 2) and four $\pi/2$ rotations.
Because the gate implementation time is determined primarily by the number of conditional two-qubit gates this implementation is faster by a factor of 2 than the traditional implementation with six CNOT gates (see Appendix \ref{Appendix:notation}).

We note that a similar circuit for the Toffoli gate with three global conditional operations has been proposed by Ref.~\cite{Nebendahl2009} who have used five global single-qubit $\pi/2$ and $\pi/4$ rotations and three local phase gates rather than the four local $\pi/2$ rotations and the four local phase gates used here.

Due to the commutation of the operators $\sigma_k^z$ the $\Ggate{}$-gate can be represented as a product of three $\sigma_j^z\sigma_k^z$ factors,
$\Ggate{(\phi)} = e^{i \phi \sigma_1^z \sigma_2^z} e^{i\phi \sigma_2^z \sigma_3^z} e^{i \phi \sigma_1^z \sigma_3^z}$.
Because of the property
\be
J_{\beta}^{2} = \frac{N}{4}\id + \frac{1}{2}\tsum_{j<k}^{N}\sigma_{j}^{\beta}\sigma_{k}^{\beta},\label{Jz}
\ee
where $J_{\beta} = \frac{1}{2}\sum_{j=1}^{N}\sigma_{j}^{\beta}$ ($\beta=x,y,z$) is the collective spin projection operator,
the $\Ggate{}$-gate is equivalent, up to a phase factor, to a similar gate proposed by M\o lmer and S\o rensen \cite{Molmer} for creation of GHZ states \cite{Kirchmair}.
Such global gates can be implemented by a single driving field interacting uniformly with all qubits of the desired target gate.
The $J_z^2$ gate is a part of a universal set of gates proposed by Nebendahl \etal\ \cite{Nebendahl2009} as an efficient alternative of the standard toolbox involving the CNOT or the two-qubit c-phase gate.

Rather than using rotations and phase gates, all circuits from this paper can be written with Hadamard gates and T-gates, traditionally used in information theory. This can be achieved by using the relations \eqref{HtoR}. For the cc-phase gate we have (in operator form, to be read from left to right):
\be
H_1 \Ggate{(\tfrac\pi4)} H_1 T^{*}_1 H_1 \Ggate{(\tfrac\pi4)} H_1 \Ggate{(\tfrac\pi8)} T_1 T_2 T_3,
\ee
where the index indicates the qubit number.

\subsubsection{The Fredkin gate}

We have constructed a circuit for the Fredkin gate \eqref{FG} with four global two-qubit gates, shown in Fig. \ref{global-gates} (bottom).
Here qubits 2 and 3 are swapped conditionally on the state of qubit 1.
In operator form (to be read from left to right):
\begin{align}
&P_3 \T_3(\tfrac\pi2) \Ggate{(\tfrac\pi4)} P_2 P_3 \T_1(\tfrac{3\pi}{8}) \T_2(\tfrac{5\pi}{8}) \T_3(\tfrac{7\pi}{8})\Ggate{(\tfrac\pi8)} P_2 \Ggate{(\tfrac\pi4)} \times\notag\\
&\times \T_2(\tfrac{3\pi}{4}) Q_2 P_3 \T_3(\tfrac{\pi}{2}) \Ggate{(\tfrac\pi4)} P_3 \T_3(\tfrac{\pi}{4}),
\end{align}
where $Q_k$ represents a $\pi/4$ rotation on qubit $k$, $Q_k = e^{-i\frac{\pi}{8}\sigma_k^x}$.

\subsubsection{Four-qubit ccc-phase gate}

The same approach can be used to construct efficient circuits for higher-conditional gates.
Here we present an example for the four-qubit ccc-phase gate, which in operator form reads
\be
\bt{cc}
\bt{c}\includegraphics[width=0.08\columnwidth]{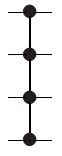}\notag\et
& \bt{c} \\ $= {\id} - 2 \ket{1111} \bra{1111} $ .  \et
\et
\ee
This gate can be obtained with the following sequence of interactions, containing 7 global $\GGgate{}$-gates (to be read from left to right):
\begin{align}
&\PG \GGgate{(\tfrac\pi4)} \TG(\tfrac\pi2) P_{123} \GGgate{(\tfrac\pi4)} \TG(-\tfrac\pi4) \PG  \T_4(-\tfrac\pi{16}) \GGgate{(\tfrac\pi4)} \PG \GGgate{(\tfrac\pi4)} \times \notag\\
&\times \T_4(-\tfrac\pi4) P_4 \T_4(\tfrac\pi8) \TG(\tfrac\pi{16}) \PG \GGgate{(\tfrac\pi4)} \TG(-\tfrac\pi4) \PG \T_4(\tfrac\pi2) \times \\
&\times \GGgate{(\tfrac\pi4)} \TG(-\tfrac\pi4) \PG \GGgate{(\tfrac\pi{16})} \T_4(\tfrac\pi2) \TG(\tfrac{9\pi}{16}), \notag
\end{align}
where
\bse
\begin{align}
P_{123} &= e^{-i \frac\pi 4 \left(\sigma_1^x+\sigma_2^x+\sigma_3^x\right)}, \\
\PG &= e^{-i \frac\pi 4 (\sigma_1^x+\sigma_2^x+\sigma_3^x+\sigma_4^x)}, \\
\TG(\phi) &= e^{-i \phi (\sigma_1^z+\sigma_2^z+\sigma_3^z+\sigma_4^z)}, \\
\GGgate{(\phi)} &= e^{i \phi (\sigma_1^z \sigma_2^z + \sigma_1^z \sigma_3^z + \sigma_1^z \sigma_4^z + \sigma_2^z \sigma_3^z + \sigma_2^z \sigma_4^z + \sigma_3^z \sigma_4^z)}.
\end{align}
\ese
Numerical simulations suggest that 7 is the minimum number of global gates $\GGgate{(\phi)}$. If we allow for a single non-global two-qubit gate, this number is reduced to 6.
Barenco \etal~ proposed a scheme with 13 conditional gates -- 6 CNOTs and 7 c-phase gates \cite{Barenco1995}, which is twice as many as we propose.
Other proposals exist, which prescribe fewer conditional gates \cite{CCCC}. However, they use either ancilla qubits or ancilla states.

\subsection{Implementation using nearest-neighbour interaction}

If the qubits interact only with their nearest neighbours, as in circuit QED, a global addressing produces the gate
\be
\Ngate{(\phi)} = e^{i \phi (\sigma_1^z \sigma_2^z + \sigma_2^z \sigma_3^z)},
\ee
rather than the $\Ggate{}$-gate used above.
Here the middle qubit 2 interacts with the outer qubits 1 and 3, while qubits 1 and 3 do not interact directly.
The $\Ngate{}$-gate produces less entanglement than the $\Ggate{}$-gate; hence the quantum circuits with $\Ngate{}$-gates are longer than with $\Ggate{}$-gates.
We point out that the $\Ngate{}$-gate, like the $\Ggate{}$-gate, can be achieved in a single interaction step by driving all qubits globally with the same field.

\subsubsection{The cc-phase gate}

\begin{figure*}
\label{NN-gates}
\bt{c}
\includegraphics[width=0.75\textwidth]{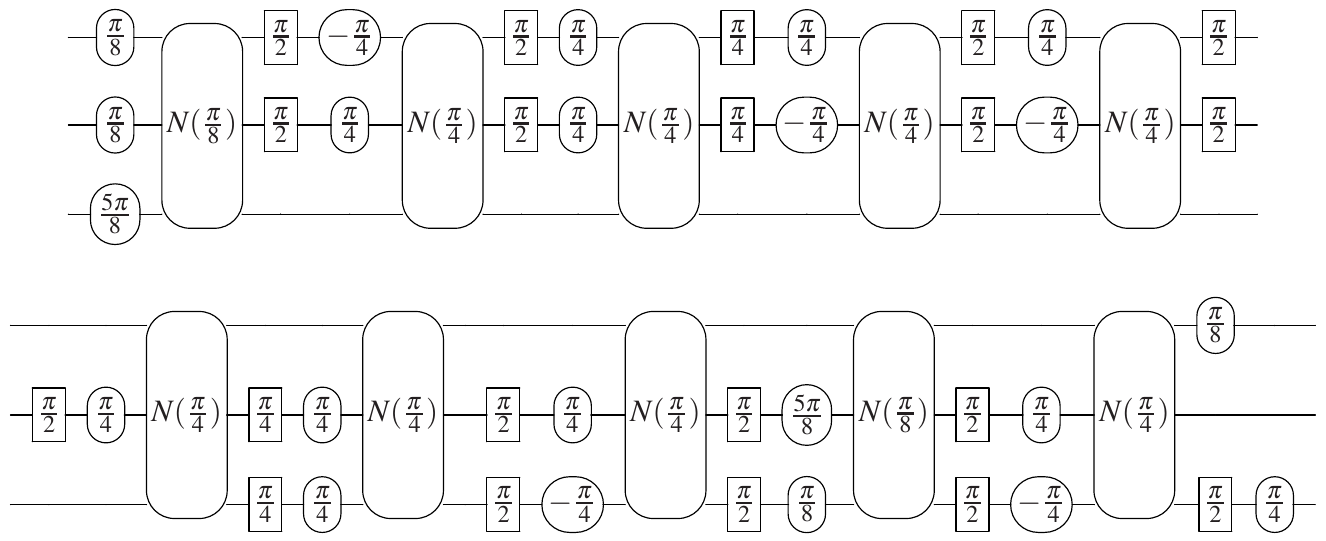}\notag
\et
\caption{Nearest-neighbour realization of the three-qubit cc-phase gate (top) and the Fredkin gate (bottom).}
\end{figure*}

The cc-phase gate can be obtained using the circuit, shown in Fig. \ref{NN-gates} (top).
Now five $\Ngate{}$-gates are necessary, as opposed to the three $\Ggate{}$-gates used by the global implementation.
We note here that the number of conditional gates is the same as in the standard circuit model  \cite{Barenco1995,Nielsen2000};
 however, the standard circuit (see Appendix \ref{Appendix:notation}) uses conditional two-qubit gates between each pair of qubits, rather than between nearest neighbors only.
Here we use only nearest-neighbors interactions.

\subsubsection{The Fredkin gate}

We have also constructed a circuit for the Fredkin gate, given in Fig. \ref{NN-gates} (bottom).
The control qubit is 1 and the target qubits to be swapped are 2 and 3. Like the cc-phase gate, the Fredkin gate demands five conditional gates.

\section{Physical realization with trapped ions}

Below we consider a few natural physical realizations of the $\Ggate{}$-operator with trapped ions. Two architectures are considered -- laser-driven traps and magnetic-gradient traps.
Because the $\Ggate{}$-operator bears the highest cost in time duration we focus our attention at its construction.
It is noteworthy that the $\Ggate{}$-operator is symmetric upon permutation of the indices of the $\sigma^z$ operators, and it involves all qubits.
The important implication from here is that, in laser-driven traps, it can be implemented in a single shot by a bichromatic pulse, i.e. by addressing all three ions simultaneously with the same field, in a global fashion. In magnetic-gradient traps, it is delivered simply by free evolution.

\subsection{Realizations with bichromatic laser interaction}

We consider a linear ion crystal of $N$ identical two-state ions with transition frequency $\omega_{0}$ confined in a Paul trap along the $z$ axis.
We assume that the ion crystal interacts with a bichromatic laser field along the transverse $x$ direction with laser frequencies $\omega_{\rm r}=\omega_{0}-\omega_{\rm{c.m.}}+\delta$ (``red'') and $\omega_{\rm b}=\omega_{0}+\omega_{\rm{c.m.}}-\delta$ (``blue'') tuned near the center-of-mass ({\rm{c.m.}}) mode $\omega_{\rm{c.m.}}$ with detuning $\delta$, ($\omega_{\rm{c.m.}}\gg |\delta|$).
The interaction Hamiltonian in the Lamb-Dicke limit $(\eta\ll 1)$ and the rotating-wave approximation is given by \cite{Haljan2005} ($\hbar=1$)
\begin{equation}
H_{\rm I} = \tsum_{k=1}^{N}g_{k}(t)\sigma(\varphi_{k}^{+})(a^{\dag}e^{i\delta t-i\varphi_{k}^{-}}+ ae^{-i\delta t+i\varphi_{k}^{-}}).\label{Hbich}
\end{equation}
Here $g_{k}(t)=\eta\Omega_{k}(t)/\sqrt{N}$ is the individual time-dependent spin-phonon coupling for ion $k$, $\eta$ is the single-ion Lamb-Dicke parameter, and $\Omega_{k}(t)$ is the Rabi frequency ``unit''. $a^{\dag}$ and $a$ are creation and annihilation operators of the {\rm{c.m.}} vibrational mode and $\sigma(\varphi_{k}^{+})=e^{-i\varphi_{k}^{+}}\sigma_{k}^{+}+e^{i\varphi_{k}^{+}}\sigma_{k}^{-}$ with $\sigma_{k}^{\pm}$ being the spin raising and lowering operators for ion $k$.
In terms of the blue and the red laser phases $\varphi_{k}^{\rm b}$ and $\varphi_{k}^{\rm r}$, the spin and motional phases in Eq.~\eqref{Hbich} are defined as
 $\varphi_{k}^{+}=\frac{1}{2}(\varphi_{k}^{\rm b}+\varphi_{k}^{\rm r})$ and $\varphi_{k}^{-}=\frac{1}{2}(\varphi_{k}^{\rm b}-\varphi_{k}^{\rm r})$, respectively.
Hereafter we assume that the blue and the red lasers are counterpropagating, which implies that $\varphi_{k}^{+}=0$ and $\varphi_{k}^{-}=\varphi_{k}$, such that $\sigma(\varphi_{k}^{+})=\sigma_{k}^{x}$.

\subsubsection{The $\sigma_{x}$ gate}

Let us consider the case of $N$ ions addressed simultaneously with a suitably detuned bichromatic field, as in the S{\o}rensen-M{\o}lmer two-qubit c-phase gate \cite{Sorensen}.
Setting $g=g_{k}(t)$ and $\varphi=\varphi_{k}$ in Eq.~\eqref{Hbich}, the propagator becomes
\begin{equation}
U = D(\alpha) \exp\left[{i\frac{4g^{2}}{\delta^{2}}(\delta t-\sin\delta t)J_{x}^{2}}\right],
\label{Uoff}
\end{equation}
where $D(\alpha)=e^{\alpha a^{\dag}-\alpha^{\dag}a}$ is the displacement operator with $\alpha=2g J_{x}e^{-i\phi}(1-e^{i\delta t})/\delta$ being the displacement parameter.
At time
\be
t_{\text{gate}} = \frac{2\pi}{\delta},
\ee
the motional displacement in Eq.~\eqref{Uoff} vanishes, $\alpha=0$, while the spin component acquires a nonzero geometric phase shift,
\be
\label{UMS}
U = \exp\left(i\frac{8\pi g^{2}}{\delta^{2}}J_{x}^{2}\right).
\ee
Using the property \eqref{Jz} we finally obtain
\be\label{propagator bichromatic}
U = \exp\left(i\frac{2\pi g^{2}N}{\delta^{2}}\right) \exp\left(i\frac{4\pi g^{2}}{\delta^{2}}\tsum_{j<k}^{N}\sigma_{j}^{x}\sigma_{k}^{x}\right).
\ee
Note that in all formulas above, $N=3$ for the three-qubit gates and $N=4$ for the four-qubit gates.
Apart from the unimportant phase factor, the propagator of Eq.~\eqref{propagator bichromatic} is exactly the $\Ggate{}$-gate of Eq.~\eqref{B} needed in our implementations,
with the relation
\be
\phi = \frac{4\pi g^{2}}{\delta^{2}}.
\ee

We note here that this implementation of the $\Ggate{}$-gate with the $\sigma^x$ couplings is equivalent to an implementation in the $\sigma^z$ basis (described in the charts above) rotated at an angle of $\pi/4$ (see details in the Appendix).

\subsubsection{The $\sigma_{z}$ gate}

The alternative scheme is based on collective Raman type interaction with frequency difference close to the vibrational frequency rather that the qubit frequency. We assume that the linear crystal of $N$ ions is uniformly addressed with two Raman beams with wave vector difference $\Delta \vec{k}$ pointing along the transverse $x$ direction and laser frequency difference $\Delta \omega_{\rm L}=\omega_{\rm c.m.}-\delta$. The Hamiltonian describing the laser-ion interaction in the Lamb-Dicke limit is given by \cite{Lee2005,Schneider2012}
\begin{equation}
H_{\rm I}=gJ_{z}(a^{\dag}e^{i\delta t}+ae^{-i\delta t}).\label{Hz}
\end{equation}
The unitary propagator corresponding to the Hamiltonian \eqref{Hz} is identical in form to \eqref{UMS} by replacing $J_x \rightarrow J_z$.

%%%%%%%%%%%%%%%%%%%%%%%%%%%%%%%%%%%%%%%%%%%%%%%%%%%%%%%%%%%%%%%%%%%%%%%%%%%%%%%%%%%%%%%%%%%%%%%%%%%%%%%%%%%%%%%%%%%%%%%
\subsection{Realization with magnetic-gradient ion traps}\label{Sec-magnetic}

In this ion-trap scheme a magnetic field is applied such that the ions experience a field gradient along the chain.
This creates a magnetic gradient induced coupling (MAGIC), which makes microwave or rf radiation
effective for addressing and coupling internal qubit states to the vibrational motion of the ions.
Thus the magnetic-gradient ion trap does not require laser light for single- and multiple-qubit quantum gates \cite{Mintert2001}.
The Hadamard gate is achieved with a microwave or rf $\pi/2$-pulse resonant with the qubit transition frequency, while
the phase gate can be obtained by choosing the phase of the driving field.
The conditional gate \eqref{B} can be obtained in a very simple fashion, as described in the rest of this section.
We follow the method proposed in Ref.~\cite{IvanovWunderlich}.

In a frame rotating with the ions' internal states and to second order in the ions vibrational motion the Hamiltonian can be written as \cite{Mintert2001}
\be
\label{magneticHam}
H = \tsum_{n=1}^N \omega_n a_n^{\dagger}a_n - \frac{1}{2}\tsum_{j<k}J_{jk}\sigma_{j}^{z}\sigma_{k}^{z},
\ee
where $a_n^{\dagger}$ and $a_n$ are respectively the creation and annihilation operators for the $n$th vibrational mode.
The first term is a sum of the energies of all $N$ vibrational states.
The second term represents long-range pairwise spin-spin coupling, whereby spin $j$ is coupled to spin $k$ with the coupling coefficient \cite{Johanning2009}
\be
J_{jk}=\frac{\left(g_{\rm e}\mu_\text{B}b\right)^2}{2}(\hes^{-1})_{jk}.
\ee
Here $\hes$ is the Hessian of the trap potential taken at the equilibrium positions of the ions, $b$ is the amplitude of the magnetic field gradient, $g_{\rm e}$ and $\mu_\text{B}$ are respectively the electron g-factor and Bohr's magneton.
This spin-spin coupling is only weakly sensitive to thermal excitation of the ion string. Thus, usually just Doppler cooling is sufficient to avoid unwanted thermal effects on the coupling constants $J_{jk}$ \cite{Wunderlich2004}.

The second term in Eq.~\eqref{magneticHam}, which describes the spin states, is decoupled from the first term, so the spin states can be described independently from the motional states.
The propagator corresponding to this spin term is
\be
\label{Ut}
U(\tau)=\exp{\left(\frac{i \tau}{2}\tsum_{j<k=1}^N J_{jk}\sigma^z_{j}\sigma^z_{k}\right)}.
\ee
It describes the free evolution of the $N$ spins (with no microwave or rf driving), which are only exposed to the magnetic field gradient for time period $\tau$.

\subsubsection{Equal couplings}

For equal couplings, $J_{jk}=J$, the propagator of Eq.~\eqref{Ut} reduces to the conditional gate $\Ggate{(\phi)}$ at time $\tau =2\phi/J$,
\be
\Ggate{(\phi)}=U(2\phi/J).
\ee
Thus the implementation of $\Ggate{(\phi)}$ requires no external driving. It simply takes place as a result of the free evolution of the spins for time $2\phi/J$.
Hence, in the magnetic-gradient traps the Toffoli gate and the Fredkin gate can be realized as sequences of microwave or rf pulses (producing the single-qubit gates), applied at certain times \cite{IvanovWunderlich}.

\subsubsection{Unequal couplings}

\begin{figure}
\label{unequal}
\bt{c}
\includegraphics[width=0.9\columnwidth]{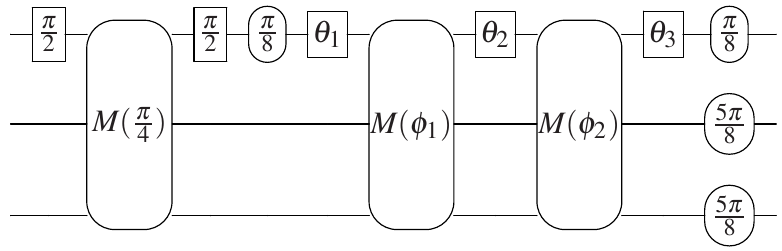}\notag
\et
\caption{Realization of the three-qubit cc-phase gate with unequal couplings $J_{jk}$. For uniform $J_{jk}$ this circuit reproduces the circuit from Fig. \ref{global-gates} (top).}
\end{figure}

Our method can be applied for different couplings $J_{jk}$, as well. In the linear Paul trap with harmonic potential, for example, where $J_{12}=J_{13}\neq J_{23}$,
the \ccc~gate is obtained with the circuit, shown in Fig. \ref{unequal}.
In operator form (to be read from left to right):
\begin{align}
P_1 \Ugate{(\tfrac\pi4)} P_1 \T_1(\tfrac\pi8) P_1(\theta_1) &\Ugate{(\phi_1)}P_1(\theta_2) \Ugate{(\phi_2)}\times \\
\times  P_1(\theta_3)&\T_1(\tfrac{\pi}8) \T_2(\tfrac{5\pi}8) \T_3(\tfrac{5\pi}8).\notag
\end{align}
Now the conditional gate \eqref{B} is defined as $\Ugate{(\phi)}=U(2\phi/J_{12})$ and $P_k(\theta)=e^{-i\frac{\theta}{2} \sigma_k^x}$. We have $\phi_1=0.375\pi$, $\phi_2=0.258\pi$, $\theta_1=1.032\pi$, $\theta_2=0.484\pi$, $\theta_3=1.484\pi$.

Note that we have relabelled the ions from $1,2,3$ to $2,1,3$. Thereby, for uniform $J_{jk}$, the above circuit is reduced to the circuit from Fig. \ref{global-gates} (top), as expected.

The corresponding circuit for the Fredkin gate is not given as it is more cumbersome.

%%%%%%%%%%%%%%%%%%%%%%%%%%%%%%%%%%%%%%%%%%%%%%%%%%%%%%%%%%%%%%%%%%%%%%%%%%%%%%%%%%%%%%%%%%%%%%%%%%%%%%%%%%%%%%%%%%%%%%%
\section{Discussion}
%%%%%%%%%%%%%%%%%%%%%%%%%%%%%%%%%%%%%%%%%%%%%%%%%%%%%%%%%%%%%%%%%%%%%%%%%%%%%%%%%%%%%%%%%%%%%%%%%%%%%%%%%%%%%%%%%%%%%%%

We note here that several previous works have paid special attention to the single-qubit phase gates involved in the construction of the Toffoli gate because in some quantum computing platforms these gates are the ones that are most difficult to construct.
As mentioned above, the standard implementation of the Toffoli gate uses six such gates in five groups (meaning that two of them can be applied simultaneously to different qubits) \cite{Barenco1995,Nielsen2000}.
These numbers have been reduced to four phase gates in two groups by using ancilla states and six CNOT gates.
Amy \etal\ \cite{Amy} reduced the T-depth to 3.
In all of our implementations proposed here we have the same numbers --- four single-qubit phase gates in two groups --- but without ancilla states and using 3 global two-qubit gates.
Hence in the information theory language, our construction of the Toffoli gate has T-depth equal to 2.
Therefore our construction of the Toffoli gate is as compact as the best T-depth of achieved previously.
However, we emphasize that the T-depth is of little interest to implementations with trapped ions because for them the number of two-qubit conditional gates (Clifford gates in information science language) is of primary concern because this number determines the speed of the total gate.

\section{Conclusions}\label{Sec-conclusions}

In this work we have proposed mathematical constructions of the Toffoli gate (or the closely related two-qubit cc-phase gate), the four-qubit ccc-phase gate and the Fredkin gate using global two-qubit gates.
The global gates employ either i) global interaction, where all pairs of qubits interact collectively, or ii) nearest-neighbour interaction, where each qubit interacts with its nearest neighbour qubits only. The Toffoli gate demands three global two-qubit gates, which is a factor of 2 faster than conventional proposals with six CNOT gates (see Appendix \ref{Appendix:notation}). The Fredkin gate uses four global conditional gates.
The four-qubit ccc-phase gate uses seven global conditional gates or six, if one of the conditional gates is not global. Proposals exist, which prescribe fewer such gates, however resorting to ancilla states or ancilla qubits. We point out that no ancillas are used in our circuits.

We also give various physical realizations.
Global conditional gates are perfectly suited for systems such as ion traps, where ions interact equally with an applied external field. Two types of traps are considered. In the standard laser-driven traps, each conditional gate is implemented using a single bichromatic pulse; two particular realizations are given. In the microwave-driven traps the conditional gates are achieved simply from the \emph{free evolution}, i.e. with no external field driving -- we only let the ions evolve freely for a predetermined time period under the influence of the magnetic-gradient field.

Nearest-neighbour interaction is suitable, e.g., for cirquit QED, where the nearest neighbour coupling arises naturally from the hopping of photons between adjacent cavities in the regime of photon blockade \cite{Illuminati}. Our circuits of the Toffoli and the Fredkin gate prescribe five nearest-neighbour conditional gates. We note that all such gates are also global in the sense that they can be achieved by driving all qubits simultaneously; therefore they can be realized in a single interaction step, too.

Finally, we point out that our numerical methods cannot give a proof that our circuits are optimal. But we note that after extensive search no better circuits using the described gate set were found.

%%%%%%%%%%%%%%%%%%%%%%%%%%%%%%%%%%%%%%%%%%%%%%%%%%%%%%%%%%%%%%%%%%%%%%%%%%%%%%%%%%%%%%%%%%%%%%%%%%%%%%%%%%%%%%%%%%%%%%%%%%%%%%%%%%%
\acknowledgments

This work has been supported by the EC Seventh Framework Programme under grant agreement No. 270843 (iQIT).

\appendix

%%%%%%%%%%%%%%%%%%%%%%%%%%%%%%%%%%%%%%%%%%%%%%%%%%%%%%%%%%%%%%%%%%%%%%%%%%%%%%%%%%%%%%%%%%%%%%%%%%%%%%%%%%%%%%%%%%%%%%%
\section{Notation for gates}\label{Appendix:notation}
%%%%%%%%%%%%%%%%%%%%%%%%%%%%%%%%%%%%%%%%%%%%%%%%%%%%%%%%%%%%%%%%%%%%%%%%%%%%%%%%%%%%%%%%%%%%%%%%%%%%%%%%%%%%%%%%%%%%%%%

The traditional implementation of the Toffoli gate uses 6 CNOT gates and 10 single-qubit gates \cite{Barenco1995,Nielsen2000},
\bc
\bt{ccc}
\bt{c}
\includegraphics[width=0.9\columnwidth]{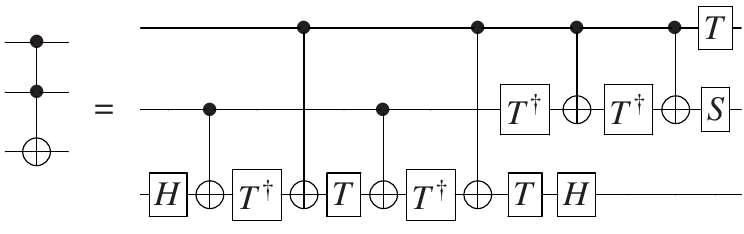}\notag
\et
\et
\ec
The three-qubit c-phase gate is constructed with the same circuit but without the Hadamard gates.
\bc
\bt{ccc}
\includegraphics[width=0.9\columnwidth]{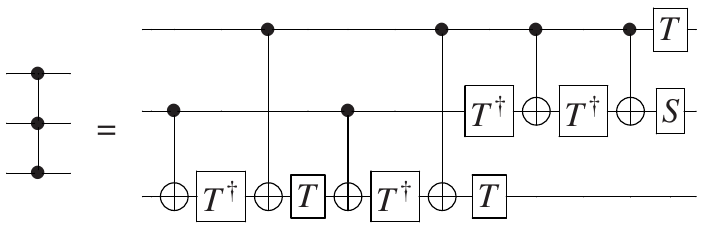}\notag
\et
\ec

The traditional single-qubit gates are
\bse
\bea
\includegraphics[width=0.13\columnwidth]{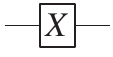}
& \left[\ba{cc} 0 & 1 \\ 1 & 0 \ea\right]
& \text{Pauli}\ \sigma^x
 \\
\includegraphics[width=0.13\columnwidth]{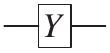}
& \left[\ba{cc} 0 & -i \\ i & 0 \ea\right]
& \text{Pauli}\ \sigma^y
 \\
\includegraphics[width=0.13\columnwidth]{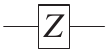}
& \left[\ba{cc} 1 & 0 \\ 0 & -1 \ea\right]
& \text{Pauli}\ \sigma^z
 \\
\includegraphics[width=0.13\columnwidth]{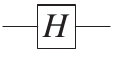}
& \tfrac{1}{\sqrt{2}} \left[\ba{cc} 1 & 1 \\ 1 & -1 \ea\right]
& \text{Hadamard}
 \\
\includegraphics[width=0.13\columnwidth]{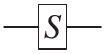}
& \left[\ba{cc} 1 & 0 \\ 0 & i \ea\right]
& \pi/4\ \text{phase gate}
 \\
\includegraphics[width=0.13\columnwidth]{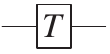}
& \left[\ba{cc} 1 & 0 \\ 0 & e^{i\pi/4}\ea\right]
& \pi/8\ \text{phase gate}
 \\
\includegraphics[width=0.13\columnwidth]{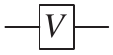}
& \tfrac{1}{\sqrt{2}} \left[\ba{cc} 1 & i \\ i & 1 \ea\right]
& \sqrt{\text{NOT}}\ \text{gate}
\eea
\ese

In a closed qubit, all unitary transformations are SU(2), i.e. with unit determinant.
In this case, we use the physically relevant gates

\begin{center}

\bt{crl}
\bt{c}\includegraphics[width=0.13\columnwidth]{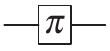}\et
&
\bt{c} $\left[\ba{rr}
0 & -i \\
-i & 0
\ea\right]$ \et
 & $\pi$ pulse, $e^{-i \frac \pi 2 \sigma^x}$
 \\ \\
\bt{c}\includegraphics[width=0.13\columnwidth]{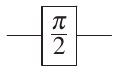}\et
&
\bt{c} $\frac{1}{\sqrt{2}}\left[\ba{rr}
1 & -i \\
-i & 1
\ea\right]$ \et
 & $\pi/2$ pulse, $P = e^{-i \frac \pi 4 \sigma^x}$
 \\ \\
\bt{c} \includegraphics[width=0.13\columnwidth]{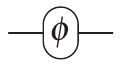} \et
&
\bt{c} $\left[\ba{cc}
e^{-i\phi} & 0 \\
0 & e^{i\phi}
\ea\right]$ \et
 & phase gate, $\T{(\phi)} = e^{-i \phi \sigma^z}$
\et

\end{center}

Unlike the preceding gates, these gates can be implemented in a closed qubit, without ancilla states.
Obviously,
\begin{align}
\label{HtoR}
\sigma^x &= i e^{-i \frac \pi 2 \sigma^x}, \\
S &= e^{i\pi/4} \T{(\pi/4)}, \\
T &= e^{i\pi/8} \T{(\pi/8)}, \\
H &= i \T{(\pi/4)} e^{-i \frac \pi 4 \sigma^x} \T{(\pi/4)}.
\end{align}

Useful relations between $x$ and $z$ bases:
\be
H e^{-i \alpha \sigma^x} H = e^{-i \alpha \sigma^z} = \left[\ba{cc} e^{-i \alpha} & 0 \\ 0 & e^{i \alpha} \ea\right]
\ee
\be
H e^{-i \alpha \sigma^z} H = e^{-i \alpha \sigma^x} = \left[\ba{cc} \cos \alpha & -i\sin\alpha \\ -i\sin\alpha & \cos \alpha \ea\right]
\ee
\be
(H_1\otimes H_2) e^{-i \alpha \sigma_1^x\otimes \sigma_2^x} (H_1\otimes H_2) = e^{-i \alpha \sigma_1^z \otimes \sigma_2^z}
\ee
\be
(H_1\otimes H_2) e^{-i \alpha \sigma_1^z\otimes \sigma_2^z} (H_1\otimes H_2) = e^{-i \alpha \sigma_1^x \otimes \sigma_2^x}
\ee

%%%%%%%%%%%%%%%%%%%%%%%%%%%%%%%%%%%%%%%%%%%%%%%%%%%%%%%%%%%%%%%%%%%%%%%%%%%%%%%%%%%%%%%%%%%%%%%%%%%%%%%%%%%%%%%%%%%%%%%
\section{Numerical procedure}\label{Appendix:numerics}
%%%%%%%%%%%%%%%%%%%%%%%%%%%%%%%%%%%%%%%%%%%%%%%%%%%%%%%%%%%%%%%%%%%%%%%%%%%%%%%%%%%%%%%%%%%%%%%%%%%%%%%%%%%%%%%%%%%%%%%
We follow the procedure as described in Ref. \cite{IvanovWunderlich}. First, we form a generic circuit containing $N_\text{G}$
two-qubit gates $\Bgate{}$:
\be
\tilde{F}=\Bgate{(\phi_{N_\text{G}})} ~L_{N_\text{G}} \ldots ~\Bgate{(\phi_3)} ~L_3 ~\Bgate{(\phi_2)} ~L_2 ~\Bgate{(\phi_1)} ~L_1,
\ee
where $L_k$ implements a generic local operation to each qubit, $L_k\in \text{SU}(2)\otimes \text{SU}(2)\otimes \text{SU}(2)$,
and $\Bgate{}$ represents one of the conditional gates $\Ggate{}$, $\Ngate{}$ or $\Ugate{}$.
An element $h$ from SU(2) can be constructed, up to an unimportant global phase, in the following way:
\be
h=\T(\varphi_1)\exp(-\tfrac{i}{2}\theta\sigma_x)\T(\varphi_2).
\ee
This represents a pulse with area $\theta$, surrounded by two phase gates. Because the phase gates $\T(\varphi)$ commute with the operator $\Bgate{}$,
$\T(\varphi_1)$ from $L_k$ can be combined with $\T(\varphi_2)$ from $L_{k+1}$ into a single phase gate.
Therefore, without loss of generality, we take all $\varphi_2$ to be 0.

Then we proceed with a numerical minimization of the distance
\be
D=\sum_{i,j}\left|F_{ij}-\tilde{F_{ij}}\right|,
\ee
with $F$ being the target gate.
We use Newton's gradient-based method to determine the variables $\phi_k$, $\varphi_k$ and $\theta_k$, which yield $D=0$.
From all solutions obtained, we choose those with the fewer number of single-qubit gates.
To minimize the number of the two-qubit gates $\Bgate{(\phi)}$, we start from a small number $N_\text{G}$, which we gradually increase, until we reach a solution to $D=0$. Because we use a local optimization algorithm, we iteratively pick the initial values of the variables using a Monte-Carlo scheme.

%%%%%%%%%%%%%%%%%%%%%%%%%%%%%%%%%%%%%%%%%%%%%%%%%%%%%%%%%%%%%%%%%%%%%%%%%%%%%%%%%%%%%%%%%%%%%%%%%%

\end{document}